# Compound plasmonic vortex generation based on spiral nanoslits

Changda Zhou, Zhen Mou, Rui Bao, Zhong Li and Shuyun Teng*

*Shandong Provincial Key Laboratory of Optics and Photonic Device & Shandong Provincial Engineering and Technical Center of Light Manipulations, School of Physics and Electronics, Shandong Normal University, Jinan 250014, China*
\*Correponding author e-mail:tengshuyun@sdnu.edu.cn

In view of wide applications of structured light fields and plasmonic vortices, we propose the concept of compound plasmonic vortex and design several structured plasmonic vortex generators. This kind of structured plasmonic vortex generators consists of multiple spiral nanoslits and they can generate two or more concentric plasmonic vortices. Different from Laguerre-Gaussian beam, the topological charge of the plasmonic vortex in different region is different. Theoretical analysis lays the basis for the design of radially structured plasmonic vortex generators and numerical simulations for several examples confirm the effectiveness of the design principle. The discussions about the interference of vortex fields definite the generation condition for the structured vortex. This work provides a design methodology for generating new vortices using spiral nanoslits and the advanced radially structured plasmonic vortices is helpful for broadening the applications of vortex fields.

## 1. Introduction

Structured light usually refers to the custom light field with nonuniform intensity, phase or polarization distribution in space, and recently, it has attracted much attention because of its powerful impact on chemistry, biology, physics and the cross disciplinary [1]. Optical vortex with phase singularity and rotational flow is an ongoing concern subject among the studies about structured light. Generally, optical vortex can be expressed by the term of $\exp(jl\varphi)$, where $\varphi$ is the azimuthal angle and $l$ denotes the topological charge [2]. Optical vortex can provide the orbital angular momentum for light and matter interaction and its applications appear in many fields including optical communication [3], particles trapping [4] and quantum information processing [5].

Plasmonic vortex, as the high local optical vortex, comes from the superposition of surface plasmon polaritons (SPPs) excited by the nanometer scatterers and located at the interface of metal and dielectric. With comparison to the far-field vortex, plasmonic vortex realizes the in-plane control of the light field and it enables nano focusing and enhanced light-atom interaction because of the unique features of SPPs including high localization, short wavelength and extraordinary optical transmission [6-8]. Till now, several methods have been advanced to generate plasmonic vortex and they include single spiral nanoslit [9-13], spiral nanoslit with multiple arms [14-17], and distributed nanoholes [18-20]. Spiral slit has the advantages of simple structure, convenient manufacture and easy operation, and it is often used to generate low order optical vortex [21,22]. Spiral nanoslit with multiple arms is often used to generate high order optical vortex, where the small gap between the start to the end of each spiral can mitigate the influence of the SPP decay. Distributed nanoholes can be flexibly rotated to introduce the additional phase, and thus, the illumination condition is not limited into the circular polarization illumination like the former cases.

Here, we propose the compound plasmonic vortex (CPV), which consists of two or more vortices distributing in the radial regions like the case of Laguerre-Gaussian (LG) beam. This kind of beams has the advantage of parallel output, which benefits the applications of CPV in optical micromanipulation and optical integration. Section 2 gives the analysis for the formation mechanism of simple plasmonic vortex generated by the spiral nanoslits. The prototype of CPV emerges in the plasmonic vortex generated by the spiral nanoslit with multiple arms and this triggers our inspiration to generate the CPV. Section 3 presents several kinds of CPV generators and explores the distribution rules of the generated CPVs. The generation condition of CPV is discussed in section 4. We believe the advancement of CPV will be helpful for expanding the applications of compound plasmonic vortices in optical micromanipulation, dense information coding, optical sensing and quantum information processing.

## 2. Simple plasmonic vortex

One plasmonic vortex has the annular intensity distribution and its phase changes uniformly the times of $2\pi$ around the center. The nanoslit with spiral trajectory like Archimedes spiral or Fermat's spiral is often used to generate plasmonic vortex. These common spirals can be unified by $\alpha$ spiral and the trajectory of $\alpha$ spiral with the geometric charge of $l$ can be written into the following form [13],

$$r^{\alpha} = r_0^{\alpha} + \alpha r_0^{\alpha-1} \frac{l\lambda_{spp}}{2\pi}\theta \tag{1}$$

where $r_0$ is the initial radius of the spiral, $\theta$ is the azimuthal angle, $\lambda_{spp}$ is the wavelength of surface plasmonic polariton (SPP), and $l\lambda_{spp}$ denotes the gap of spiral. $\alpha$ is a real number between 1 and 2, and the spiral is the Archimedes spiral as $\alpha=1$ and the Fermat's spiral as $\alpha=2$. Theoretically, an Archimedes spiral with the geometric charge of $l$ can generate the plasmonic vortex with the topological charge of $l\pm 1$ under the circular polarization light illumination and the amplitude of vortex is proportional to Bessel function of $l\pm 1$ order [23], where the plus and minus signs correspond to the left- and right-handed circularly polarization (LCP and RCP) light illumination. In fact, only low order plasmonic vortex can be generated by Archimedes spiral, and for the spiral with a certain $\alpha$, the highest order of the generated intact plasmonic vortex has a valid region [13]. In order to obtain the higher order plasmonic vortex, the spiral slit with multiple arm is usually used, and the trajectory of the spiral slit with multiple arms can be expressed by [14],

$$r = r_0 + \frac{\lambda_{spp}}{2\pi}\mathrm{mod}(l\theta, 2\pi) \tag{2}$$

where mod($a$, $b$) represents the remainder of the division of $a$ by $b$.

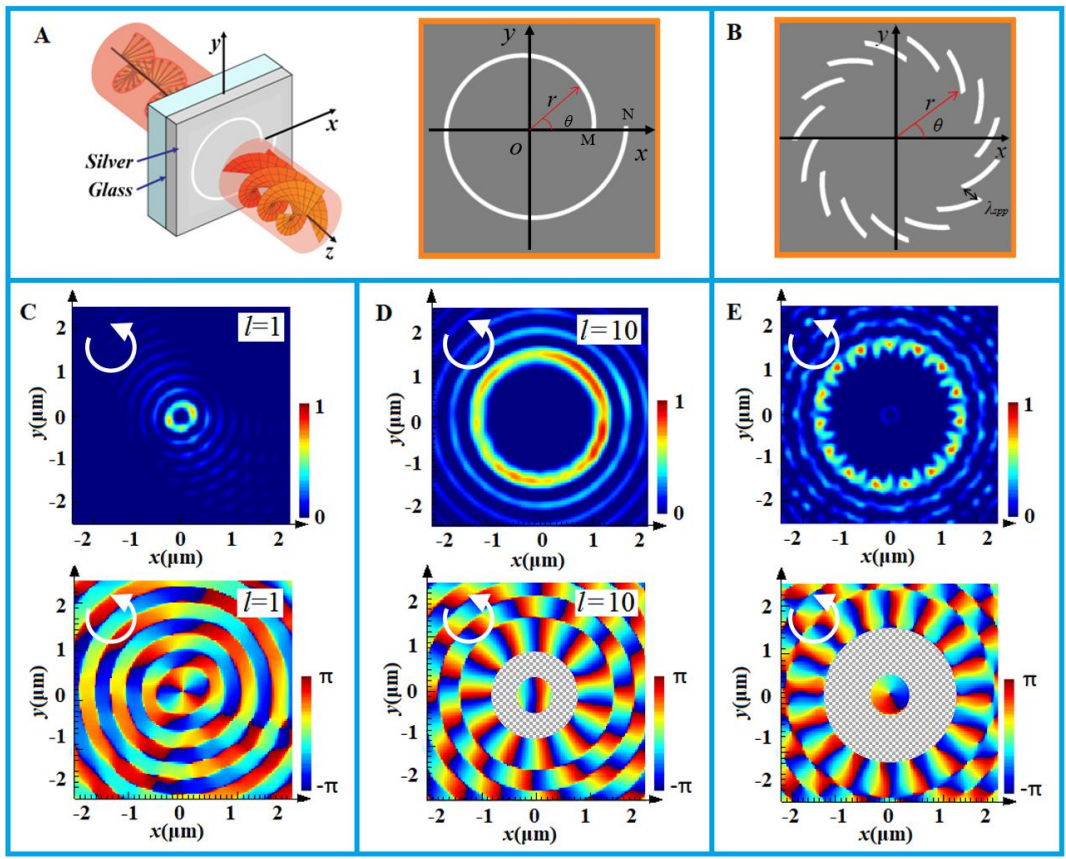

Fig.1 A: Schematic diagram of plasmonic vortex generation by a spiral, B: the structure of a spiral with many arms, C: the diffraction intensity and phase distribution of Archimedes spiral with $l=1$ and $r_0=4\lambda_{spp}$, D: the diffraction intensity and phase distribution of $\alpha$ spiral with $l=10$, $r_0=4\lambda_{spp}$ and $\alpha=1.2$, E: the diffraction intensity and phase distributions of Archimedes spiral with 15 arms, $r_0=6\lambda_{spp}$ and $l=15$. The white arrows in C, D and E denote the incident LCP.

Fig.1A shows the schematic diagram for the formation of plasmonic vortex by a spiral slit, and the separation of the start point $M$ and the end point $N$ shown in the right structure denotes the gap of spiral. Fig.1B shows the the structure of a spiral slit with many arms, the separation from the end of the former slit segment to the start of the next slit segment equals to $\lambda_{spp}$. Fig.1C gives the intensity and phase distributions of an Archimedes spiral slit etched in the silver film with the thickness of 150nm, where the width of slit takes 150nm, and the parameters of $l$ and $r_0$ take 1 and $4\lambda_{spp}$, respectively. It is easy to see that the intensity distribution is a bright ring and the phase changes twice of $2\pi$ in anticlockwise direction. This indicates the plasmonic vortex of order 2 is generated.

Fig.1D gives the intensity and phase distributions of an $\alpha$ spiral slit with $\alpha=1.2$ and $r_0=4\lambda_{spp}$, where the geometric charge of spiral takes 10. One can see that the intensity distribution has a larger bright ring and the phase changes 11 times of $2\pi$ in anticlockwise direction. This indicates the plasmonic vortex of order 11 is generated. Fig.1E gives the intensity and phase distributions of the spiral slit with 15 arms and $r_0=6\lambda_{spp}$, where the geometric charge of each spiral segment takes 15. One can see that the intensity distribution consists of many bright spots arranged on a larger circle and one small weak ring. The phase distribution contains two characteristic regions. The phase at outer region changes 16 times of $2\pi$ in anticlockwise direction and the one at inner region changes $2\pi$ in anticlockwise direction. For convenience observation, we shield the transition region between two characteristic phase regions by a gray panel. Obviously, these intensity and phase distributions are different from the results in Fig. 1D, where the phase at the inner region is uniform. However, in usual studies, the inner diffraction rule of the spiral slit with many arms is neglected by the researchers because of the weak intensity. Actually, it is a passively generated CPV, which consists of a deuterogenic plasmonic vortex of order 1 [24] and a plasmonic vortex of order 16.

It needs to be pointed out that the simulated results in Fig. 1 are obtained by using the finite difference time domain method [25, 26]. The SPP wavelength equals to $\lambda_{spp}=613$nm for the illuminating wavelength of 633nm and the polarization state is the LCP. The complex dielectric constant of silver is $\varepsilon=-15.92-j1.075$, which is taken from the value given by E. D. Palik [27]. The perfectly matched layer is chosen as the adsorbing boundary and the minimum mesh cell takes 2nm. The calculation region is set 25μm×25μm×6μm and the monitor plane is set at 200nm above the silver film. Similarly, in the following content, we design the CPV generators and verify their performance still using this method.

## 3. Compound plasmonic vortex

Similar to the structure characteristic of LG beam with many characteristic regions, we expect to generate the CPV with two or more vortices and the topological charges of the vortices at different radial regions may take different values. We first design a kind of CPV generator based on two sets of $\alpha$ spirals with the geometric charges of two spirals taking $l_1$ and $l_2$, which can form a CPV consisting of two plasmonic vortices. Fig. 2 gives three CPV generators, where the first consists of a circle and a spiral, the second and the third consist of a small spiral and a larger spiral. The geometric charge of the spiral for the first CPV generator is positive. The geometric charge for the second one takes the negative value for the inner small spiral and positive value for the outer spiral. For the third one, the geometric charges of two spirals are positive. Here, the value of $\alpha$ takes1.2, the initial radius of the larger spiral takes $10\lambda_{spp}$, and the radius of circle and the initial radius of the smaller spiral take $\lambda_{SPP}$. The radii of the inner and outer circle or spiral slits are respectively represented by $r_1$ and $r_2$, as shown in Fig. 2. In order to ensure the comparative intensities of two vortices, the outer nanoslit needs to be wider than the inner nanoslit, and for above three cases, the slit width takes 220nm for the outer one and 150nm for the inner one. The diffraction intensity and phase distributions are obtained under the LCP light illumination. For convenience, we also shield the transition regions of the phase distributions by a gray panels.

Fig. 2A gives the diffraction intensity and phase distributions of a CPV generator consisting of a spiral slit with geometric charges taking 13 and a circular nanoslit equivalent to a spiral $l=0$. One can see two uniform bright rings with the radii taking 1.5μm and 0.18μm. The outer phase changes 14 times of $2\pi$ and the inner phase changes $2\pi$ in anticlockwise direction. These results show a 14 order plasmonic vortex and a 1 order vortex generate simultaneously at different radii. Figs. 2B and 2C give the diffraction intensity and phase distributions of two CPV generators consisting of two spiral slits with different geometric charges, where the geometric charge of the large spiral takes 13, the geometric charge of the small spiral takes -1 for Fig. 2B and 1 for Fig. 2C. The results in Fig. 2B show that a bright spot appears at the center, and many almost continuous bright sports take on the outside circular trajectory. The phase at the center is uniform and the phase at the outside changes 14 times of $2\pi$ in anticlockwise direction. This indicates that a CPV consisting of a 14 order plasmonic vortex and a zero order vortex or bright spot generate at different radial positions. From Fig. 2C, one can see that o bright ring appears at the radius of 0.31μm and many bright sports takes on the outside circular trajectory with the radius taking 1.5μm. The phase at the center changes twice of $2\pi$ and the outer phase changes 14 times of $2\pi$ in anticlockwise direction. This indicates that a CPV consisting of a 14 order plasmonic vortex and a 2 order vortex generates at different radial positions.

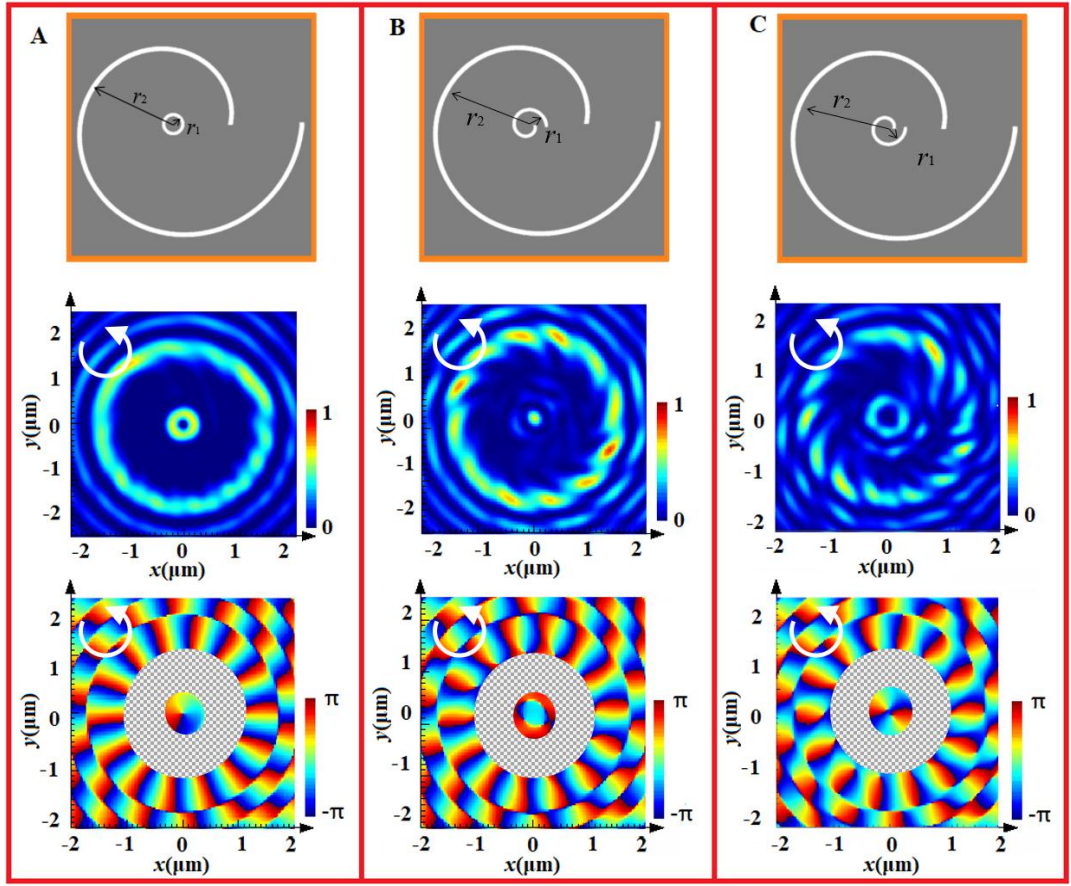

**Fig. 2** Structures of CPV generators consisting of $\alpha$ spirals and the diffraction intensity and phase distributions of the generated CPVs with A: $l_1=0$ and $l_2=13$, B: $l_1=-1$ and $l_2=13$, C: $l_1=1$ and $l_2=13$. The white arrows inserted in the patterns denote the incident LCP.

With comparison to Fig. 1B, the structures shown in Fig. 2 can actively generate the CPVs. However, these structures are difficult to generate the CPV consisting of many higher order plasmonic vortices because the highest order of the generated intact plasmonic vortex has a valid region. Thus, we use the spirals with many arms to generate the CPV consisting of two or more high order plasmonic vortices. Fig. 3 shows the schematic diagrams of this kind of CPV generators, where two or three spirals with different geometric charges are used. Fig. 3A shows the CPV generator consisting of a circle with the radius of $\lambda_{spp}$ and a spiral with 15 arms with the geometric charge taking 15 and the initial radii of spirals taking $5\lambda_{spp}$. Fig. 3B shows the CPV generator consisting of two spirals with 15 arms and 36 arms, where their geometric charges take -15 and 36, the initial radius of the inner spiral takes $5\lambda_{spp}$ and the initial radius of the outer spiral takes $12\lambda_{spp}$. Fig. 3C shows the CPV generator consisting of a circle with the radius of $\lambda_{spp}$ and two spirals with 15 arms and 36 arms, where the geometric charges take -15 and 36, the radius of the circle takes $\lambda_{spp}$, the initial radius of inner spiral takes $5\lambda_{spp}$ and the initial radius of outer spiral takes $12\lambda_{spp}$. For clearness, the radii of the circle or spiral slits from inside to outside are represented by $r_1$, $r_2$ and $r_3$, as shown in schematic diagrams of Figs. 3A-3C.

From Fig. 3A, one can see that one bright ring appears at the radius of 0.18μm and many bright sports takes on at a large circular trajectory with the radius of 1.72μm. The phase at the center changes twice of $2\pi$ and the phase at the outside changes 16 times of $2\pi$ in anticlockwise direction. This indicates that a CPV consisting of a 16 order plasmonic vortex and a 1 order vortex generates at different radial positions. This is similar to the phase distribution in Fig. 1E. From Fig. 3B, one can see that many bright spots take on at two concentric circular trajectories with the radii taking 1.55μm and 3.76μm. The phase at the center changes of $2\pi$ in anticlockwise direction, the phase in the middle changes 14 times of $2\pi$ in clockwise direction and the phase at the outside changes 37 times of $2\pi$ in anticlockwise direction. This indicates that a CPV consisting of a 1-order vortex, a 14-order vortex and a 37-order vortex generates. Since the intensity of the inner vortex is lower, it is unusable in practice. In order to increase the intensity of the inner vortex, we can actively excite it by using a circular nanoslit, like the case of Fig. 3C. With the help of the active excitation of the inner vortex, we can see clearly the bright ring appears at the center, and the phase distribution is the same as the result of Fig. 3B.

As for the physical origin of one order vortex passively generated by the spiral with many arms, as shown in Figs. 1C and 3B, it can be explained from the interference of light fields. Each segmented spiral slit has the same radial distribution and the plasmonic field excited by each spiral slit has the same contribution to the field at the center. Two slit units at the adjacent spiral slits have a certain angle difference, whose value depends on the number of arms and whose phase difference depends on the incident circular polarization. The superposition of fields coming from these slits units with the same angle difference forms 1-order Bessel field carrying a spiral phase with the topological charge of ±1. Certainly, the more the arms are, the larger the intensity of 1-order vortex is. Moreover, any order Bessel light field fluctuates with increase of the radial coordinate. This fluctuation effect influences the continuity of annular intensity distributions of the generated CPVs. The discontinuous bright spots among the intensity distributions of higher order plasmonic vortices, as shown in Figs.1-3, are just the results of the fluctuation effect of the Bessel fields.

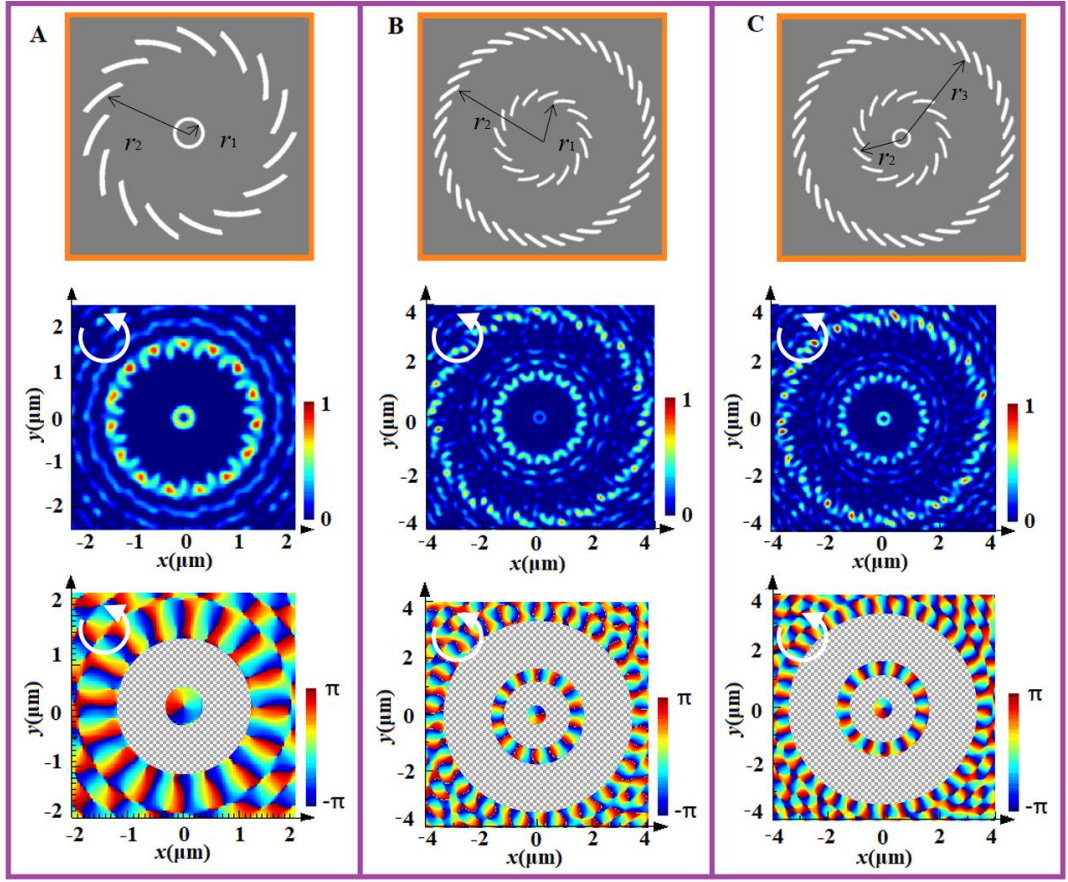

Fig. 3 Structures of CPV generators consisting of multiple-armed spirals and the diffraction intensity and phase distributions, where A: $l_1=0$ and $l_2=15$, B: $l_1=-15$ and $l_2=36$, C: $l_1=0$, $l_2=-15$ and $l_3=36$. The inserted white arrows denote the incident LCP.

## 4. Discussions

It is need to be pointed that the geometric charges of different spiral structures among the proposed CPV generators should have larger difference and their initial radii should be taken appropriate values to avoid the interference of the structure on the generated plasmonic vortex. Otherwise, as the geometric charges of two vortices are close and their initial radii are adjacent, the interference of SPP fields makes the structured vortex damaged or even invisible. In fact, as the geometric charges of two spirals are close, the amplitudes of two vortex fields are comparative. The total SPP field excited by two spirals can be written approximately as $A[\exp(jm_1\varphi)+\exp(jm_2\varphi)]$, where A denotes the amplitude of total field, $m_1$ and $m_2$ are the topological charges of two vortices and they satisfy the relation of $m_i=l_i+1$ for the LCP light illumination with $l_i$ denoting the geometric charge of spiral. Through simplifying, the total SPP field can be rewritten in the following form,

$$E = A\cos\left(\frac{m_1-m_2}{2}\varphi\right)\exp\left(j\frac{m_1+m_2}{2}\varphi\right) \quad (3)$$

As $m_1+m_2=0$, it changes into the cosine function of the position angle. As $m_1+m_2\neq0$, it must carry a spiral phase. Fig.4 shows the structures consisting of two sets of spiral nanoslits with geometric charges having smaller difference and their diffraction distributions. Fig. 4A shows the structure consisting of a circular slit with the radius of 1.1μm and the width of 120nm, and two segments of spiral nanoslits with the initial radius of 2.3μm, the geometric charge of -2 and the width of 150nm. Fig. 4B shows the structure consisting of a spiral nanoslit with the geometric charge of 1, the radius of 1.1μm and the width of 120nm, and three segments of spiral nanoslits with the initial radius of 2.3μm, the geometric charge of -3 and the width of 150nm. Fig. 4C shows the structure consisting of a circular slit with the radius of 1.1μm and the width of 120nm, and four segments of spiral nanoslits with the initial radius of 2.3μm, the geometric charge of -4 and the width of 150nm. The radii of the inner and outer circle or spiral slits are respectively represented by $r_1$ and $r_2$.

The inner and outer structures of Fig. 4A generate the plasmonic vortices with the topological charges of 1 and -1, and thus, $m_1+m_2=0$ and $m_1-m_2=2$. The intensity distribution of Fig. 4A shows that two bright spots almost appear at two symmetric positions though it seems to rotate a small angle with respect to $x$ axis. The phase distribution of Fig. 4A shows the left and right parts have the phase difference of π. They tally with the distribution of $\cos^2(\alpha)$ and abide by the rule of Eq. (3). For the inner and outer structures of Fig. 4B, the topological charges of the generated plasmonic vortices equal to 2 and -2, and thus, $m_1+m_2=0$ and $m_1-m_2=4$. The intensity distribution of Fig. 4B shows that four bright spots appear almost at $x$ and $y$ axis. The phase distribution of Fig. 4B shows the left and right parts are in phase and the up and down parts are in phase, but they have the phase difference of π. They tally with the distribution of $\cos^2(2\alpha)$ and abide by the rule of Eq. (3).

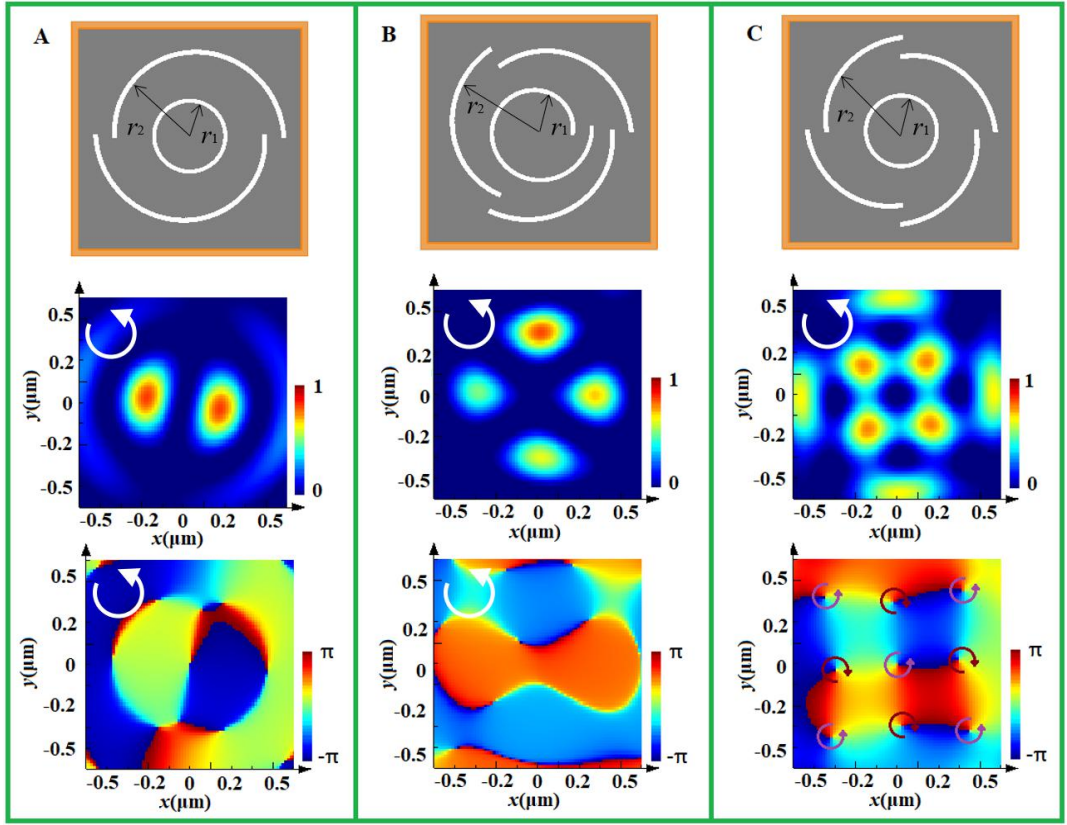

**Fig.4** Three spiral structures with the geometric charges having small difference and their intensity and phase distributions, where geometric charges of two vortices are A: $l_1=0$ and $l_2=-2$, B: $l_1=1$ and $l_2=-3$, C: $l_1=0$ and $l_2=-4$. The inserted white arrows denote the incident LCP.

For above two cases, the amplitudes of two vortices with opposite topological charge are the same. For the structures of Fig. 4C, where the topological charges of the generated plasmonic vortices are 1 and -3, $m_1+m_2=-2$ and $m_1-m_2=4$. According to Eq. (3), the intensity distribution should tally with the distribution of $\cos^2(2\alpha)$ and a vortex with the topological charge of -1 appears in the phase distribution. Fig. 4C shows that four bright spots on the outside appear almost at $x$ and $y$ axis, and they tally with the distribution rule of $\cos^2(2\alpha)$. However, there are four bright spots on the inside and they seem to tally with the distribution rule of $\sin^2(2\alpha)$. The phase distribution of Fig. 4C shows that many vortices with the topological charge of ±1 appear and the signs of the topological charges are denoted by the directions of arrows inserted in the patterns. The reason for these complex distributions is that the amplitudes of two vortices have large difference and Eq. (3) is not suitable for their superposition. Anyway, the small difference of geometric charges makes the vortices mutually interfere and the close trajectories for spiral nanoslits also influence the structured vortex distribution.

As we know, the topological charge of the vortex generated by the spiral nanoslit corresponds to the order of Bessel beam. The radius of the bright ring for high order Bessel beam is larger and it is proportional to the square root of the geometric charge of spiral. In terms of this dependent relation of the topological charge of vortex and the order of Bessel beam, the ideal CPV generator should have the following structure. The spiral nanoslit for high order vortex generation should be placed on the outside and the structure for lower order vortex generation should be placed on the inside. Moreover, the inner spiral slit trajectory should be smaller than the radius of bright ring of high order Bessel function generated by the outer spiral slits, like the cases shown in the sections 2 and 3. Because of the fluctuation characteristics of Bessel functions, the high order Bessel function on the outside of the pattern is modified by the low order Bessel function on the inside of the pattern. All these rules take on among the results shown in section 2 and section 3.

In this work, we mainly carry out theoretical analysis and numerical simulations. The preparation for the practical metasurface samples can be divided into two steps, like the former work [26, 28]. First, a silver film with the certain thickness is deposited on the glass substrate by using the magnetron sputtering method. Then, the CPV generators are fabricated by means of the focused ion beam etching method.

## 5. Conclusions

In this paper, we advance the concept of CPV with different vortex distributing at different radial region. Noticing the CPV generated passively by the spiral with multiple arms, we design the CPV generators to actively generate CPVs based on $\alpha$ spiral nanoslits. With the help of the combination of circular slit and one or more $\alpha$ spiral slits, we obtain the CPVs consisting of two or more plasmonic vortices. In order to increase the order of CPV, we combine many sets of spiral nanoslits with multiple arms to design the CPV generator and realize the higher order CPVs. The performance of the proposed CPV generators are proved by the numerical simulations. Working conditions for the CPV generators are provided and they include enough separation between spiral trajectories for different plasmonic vortices, larger topological charge difference of inner and outer plasmonic vortices, and the comparative intensities of different plasmonic vortices. This work provides a new design methodology for generating structured vortex using spiral nanoslits and the proposed CPVs may expand the applications of vortex fields.

**Acknowledgments** This work was supported by National Natural Science Foundation of China (NSFC) (10874105, 11704231).

**Disclosures** The authors declare no conflicts of interest.